\documentclass[aps,prl,twocolumn,groupedaddress,amsmath,amssymb]{revtex4}
\usepackage{makecell}
\usepackage{bbm}
\usepackage{mathrsfs}
\usepackage{subfigure}

\usepackage{algcompatible} 

\usepackage{lipsum} 
\usepackage{amsfonts}
\usepackage{color}
\usepackage{threeparttable}
\usepackage{multirow}
\usepackage{graphicx}
\usepackage{float}
\usepackage[normalem]{ulem}
\def\newblock{\hskip .11em plus .33em minus .07em}

\newcommand{\be}{\begin{equation}}
\newcommand{\ee}{\end{equation}}
\newcommand{\ba}{\begin{eqnarray}}
\newcommand{\ea}{\end{eqnarray}}

\begin{document}
	\title{ United v.s. Divided, Deconfinement of Social Tension \\
	as a Topological Phase Transition}
	\author{Chen Huang$^{1}$}
	\thanks{These two authors contributed equally.}
	\author{Jun Wu$^{1}$}
	\thanks{These two authors contributed equally.}
	\author{Xiangjun Xing$^{1,2,3}$}
	\email{xxing@sjtu.edu.cn}
	\affiliation{$^1$ Wilczek Quantum Center, School of Physics and Astronomy, Shanghai Jiao Tong University, Shanghai, 200240 China\\
	$^2$ T.D. Lee Institute, Shanghai Jiao Tong University, Shanghai, 200240 China \\
	$^3$ Shanghai Research Center for Quantum Sciences, Shanghai 201315 China}
	
	
\date{\today} 
	
\begin{abstract} 
The proverbs ``the enemy of my enemy is my friend" and alike capture the essence of many body correlations in social relations, whose violation leads to social tension.  We study how rule-breakers, who disrespect these norms, affect the structure and dynamics of signed social networks which tries to minimize social tension.  We find two dynamic phases. A friendly society exhibits a ``united phase'' where insertion of a rule-breaker only leads to localized rearrangement.  A hostile society exhibits a ``divided phase'', where insertion leads to macroscopic reorganization of social relations.  In the divided phase, starting from the utopia state, where all relations are friendly, insertion of a {\em separatist}, a particular type of rule-breaker who makes friends with only half of its neighbors, leads to fragmentation, where the society breaks into many finite size, mutually antagonistic cliques.  These phenomena are described by Ising lattice gauge theory, where social tension behave as $Z_2$ topological defects, which are confined in the united phase and deconfined in the divided phase.  We further show that the connection between social dynamics and Ising lattice gauge theory is viable independently of connectivity structure of the social network.

\end{abstract}
\maketitle

In recent decades, statistical mechanics has been successfully applied to study various social phenomena~\cite{Castellano-RMP-2009}, such as opinion dynamics~\cite{Castellano-RMP-2009,Xia-Opinion-review-2011,Lorenz-Opinion-2007,Martins-Opinion-2008}, epidemic spreading~\cite{Wang-epidemic-2017,Pastor-epidemic-2001,Castellano-epidemic-2010}, evolution of cultures~\cite{Axelrod-culture-1997,Castellano-culture-2000} and languages~\cite{Abrams-language-2003}, crowd behaviors~\cite{Vicsek-crowd-1995,Strogatz-crowd-1993}, and formation of social hierarchy~\cite{Bonabeau-hierarchies-1995,Ben-Naim-hierarchies-2005}.  In these studies, it is always assumed that individuals of the society are identical and simple, and that it is the interactions between a large number of individuals that result in complex collective phenomena.  The validity of such an assumption in social sciences is however questionable.  In almost all critical moments of history, there are always one or few ``heroes'' with extraordinary attributes who made singular impacts.   It has been constantly debated whether these great people made the history, or rather it was the history that made them great~\cite{Carlyle-hero-1869,Hook-hero-1943,Segal-hero-2000,Faulkner-hero-2008,Thompson-history-below-1966}.  According to great man theory of Thomas Carlyle~\cite{Carlyle-hero-1869}, ``The History of the world is but the Biography of great men''.  

%
\vspace{-1mm}

Social relations exhibit many-body correlations, as demonstrated by the proverbs ``friend of my friend is my friend", ``enemy of my enemy is my friend", and alike.  These correlations, which are often accepted as part of human social norms,  were first formulated as \textit{structural balance} by Heider~\cite{heider1946attitudes}, who argued that violation of these norms tend to create social tension.  The concept was further developed in the study of \textit{signed social networks}~\cite{harary1953notion,cartwright1956structural,wasserman1994social,zheng2015social}, where the links connecting individuals take values $\pm 1$ indicating friendliness/antagonism or trust/distrust~\footnote{{We note that even though real social relations are better modeled as continuous variables rather than Ising-like variables, Ising-like variables are much more preferred in statistical modeling, since it greatly simplify analysis and computation, and at the same time keeps the universal, long-scale physics. }}.  Harary~\cite{harary1953notion,cartwright1956structural} proved that a signed network is balanced, i.e. all its cycles has even number of negative links, if and only if its nodes can be split into two mutually exclusive and antagonistic groups within which all neighboring nodes are  friends.  These pioneering works have triggered a large number of studies of signed social networks~\cite{newman2006modularity,leskovec2010predicting,facchetti2011computing,esmailian2015community,jung2016personalized,aref2018measuring,kirkley2019balance,girdhar2019community,axelrod1993landscape,altafini2012consensus,altafini2012dynamics,proskurnikov2017tutorial,proskurnikov2018tutorial,hummon2003some,antal2005dynamics,antal2006social,radicchi2007social,radicchi2007universality,marvel2009energy,marvel2011continuous,traag2013dynamical,macy2003polarization,gross2019rise,saeedian2019absorbing,pham2020effect,gorski2020homophily,Pham2021}.  It has been found that many real world social networks, including online social networks, international relation networks, gene regulatory networks, and co-author networks, are indeed balanced to a high degree~\cite{leskovec2010predicting,facchetti2011computing,aref2018measuring,kirkley2019balance}.  {These examples demonstrate that real social networks tend to evolve towards balance, by reducing social tension.}  To understand how this happens, simulations were performed both on complete graph and on two dimensional lattices~\cite{antal2005dynamics,antal2006social,radicchi2007social,radicchi2007universality}.  It was found that the dynamics is fast in a friendly society and slow in a hostile society.  These two qualitatively different behaviors seem to be separated by a dynamic phase transition~\cite{radicchi2007universality}.

\begin{figure}[t!]
	\centering
	\includegraphics[width=3.1in]{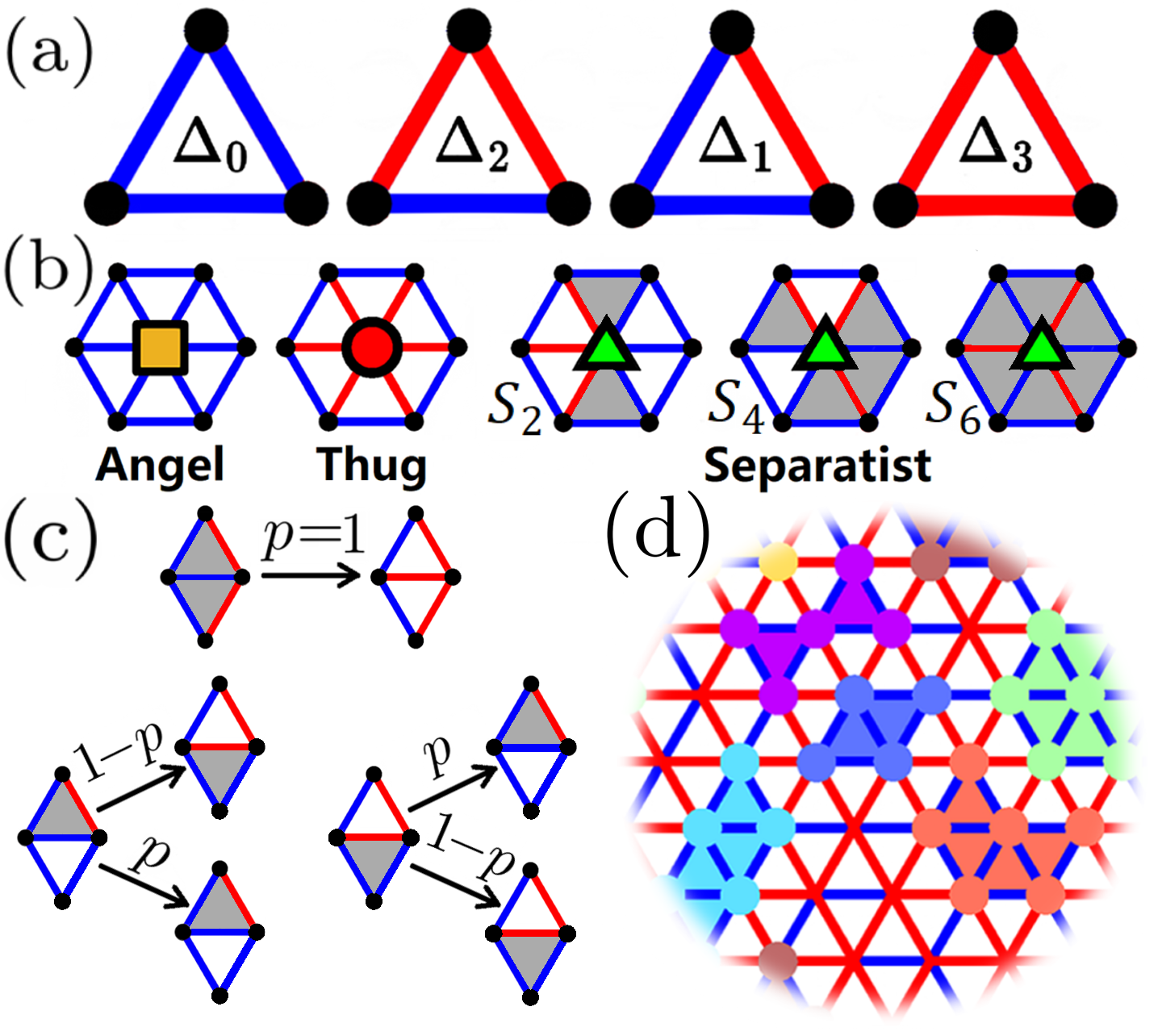}
	\vspace{-3mm}
	\caption{(a) Four possible configurations of triad. $\Delta_0$ and $\Delta_2$ are balanced, while $\Delta_1$ and $\Delta_3$ are imbalanced.  Blue link means friend, and red link means foe.  
 	(b) Rule-breakers: angel, thug and three types of separatists $S_2$, $S_4$ and $S_6$.  Shaded triads are imbalanced. (c) Link updating rules of $p$-dynamics.
	(d) A sample of fragmented society, due to insertion of a separatist in the divided phase.  Foe links percolates (red) through the entire system.  Cliques are shown in different colors.  
	}
	\label{fig::Fig1}
 \vspace{-3mm}
\end{figure}

{\bf Problem and main results} \quad Inspired by the debates between great man theory and competing theories~\cite{Hook-hero-1943,Segal-hero-2000,Faulkner-hero-2008}, in particular, the theory of ``history from below''\cite{Thompson-history-below-1966}, we would like to understand how ``rule-breakers'', who disrespect the afore-mentioned social norms, affect the balance dynamics of signed social networks.  
{There are  many attributes which distinguish ``great men'' from common people in real world.  However, from the (greatly simplified) perspective of statistical physicists, disrespect of social norm is certainly one of most outstanding feature of all great men.}  For simplicity, we first model the social network as a triangular lattice, where social relations are altered in order to minimize social tension.  We consider three types of rule-breakers, which are illustrated in Fig.~\ref{fig::Fig1}(a) and (b).  An \textit{angel} is friendly to all its neighbors, whereas a \textit{thug} is  hostile to all its neighbors, regardless of other relations in its neighborhood.  A \textit{separatist} chooses a fixed half of its neighbors as friends and the remaining half as foes.  We introduce a single rule-breaker into a randomly balanced network, and observe how the system evolves toward a new balanced state.  The resulting dyanmics turns out to depend sensitively on a ``propensity''  parameter $p$, which characterizes the likelihood of making friends if the decision does not affect the total social tension.  For $p > p_c \approx 0.64$, only local rearrangements appear, whereas for $p < p_c$, a macroscopic fraction of relations are changed before a balanced state is reached.  The duration of the re-balancing dynamics scales logarithmically with system size for $p > p_c$ and in power law for $p < p_c$.  We also insert a separatist into a {\em utopia state}, where all links are friendly.  For $p > p_c$, the network retains a system-spanning majority clique. By contrast, for $p < p_c$, the network is fragmented into many {\em cliques} of finite sizes that are connected to each other via foe links.   We call the phases with $p> p_c$ and $p < p_c$ respectively the ``united phase'' and the ``divided phase''.  We further map the social dynamics into an Ising gauge theory on triangular lattice.  Remarkably, social tension is described by topological defects in the Ising gauge model, which exhibits a topological phase transition at $p = p_c \approx  0.634$.  For $p < p_c$, defects are deconfined and diffusing freely, whereas for $p >p_c$, they are confined by long range attractions.  We further map the lattice gauge model into an Ising model on the same lattice.  Finally, we show that the mapping between signed social dynamics and Ising gauge theory is universal, independent of network structure.  


\vspace{1mm}
{\bf Model} \quad We use Ising variable $\sigma_{ij} = \pm 1$ to represent friend/foe relation between nodes $i,j$.  Three mutually connected nodes $\{i,j,k\}$ constitute a {\em triad}, which is balanced if $\sigma_{ij}\sigma_{jk}\sigma_{ki}=+1$ and imbalanced if $\sigma_{ij}\sigma_{jk}\sigma_{ki}=-1$.  As shown in Fig.~\ref{fig::Fig1}(a), there are four types of triads, $\Delta_0$ and $\Delta_2$ are balanced, whilst $\Delta_1$ and $\Delta_3$ are imbalanced.  The network is called \textit{balanced} if all its triads are balanced~\cite{heider1946attitudes,heider2013psychology,wasserman1994social,antal2005dynamics}.   We define the {\em $p$-dynamics} as follows.  At each time step we randomly pick a link $ij$, and inspect two neighboring triads.  If both triads are imbalanced, we flip the link, so that both triads become balanced; if both triads are balanced, we do nothing; if only one triad is imbalanced, we set $\sigma_{ij} = 1$ with probability $p$, and set $\sigma_{ij} = -  1$ with probability $1-p$.  Here $p$ is the {\em propensity} characterizing the friendliness of society.   As illustrated in Fig.~{\ref{fig::Fig1}}(c), whenever a link is flipped, either two neighboring imbalanced annihilate, or one imbalanced triad moves one step.  The imbalanced triads can be understood as the loci of social tension, whereas the number of imbalanced triads can be understood as the {\em total social tension}, which never increases.  The p-dynamics stops when the total social tension vanishes.

 We start from a state with each link choosing  $\pm 1$ randomly with probability $ p_0$, and turn on the $p$-dynamics.  As discussed in detail in Section I of Supplementary Information (SI), a final balanced state is always achieved, whose properties depends on $p$ but not on $ p_0$.  This state will hence be called a ``$p$-balanced society''. For $p = 1$, the  balanced state has no foe link, and will be called the {\em utopia state}.  
  
\begin{figure}[tp!]
	\centering
	\includegraphics[width=3.4in]{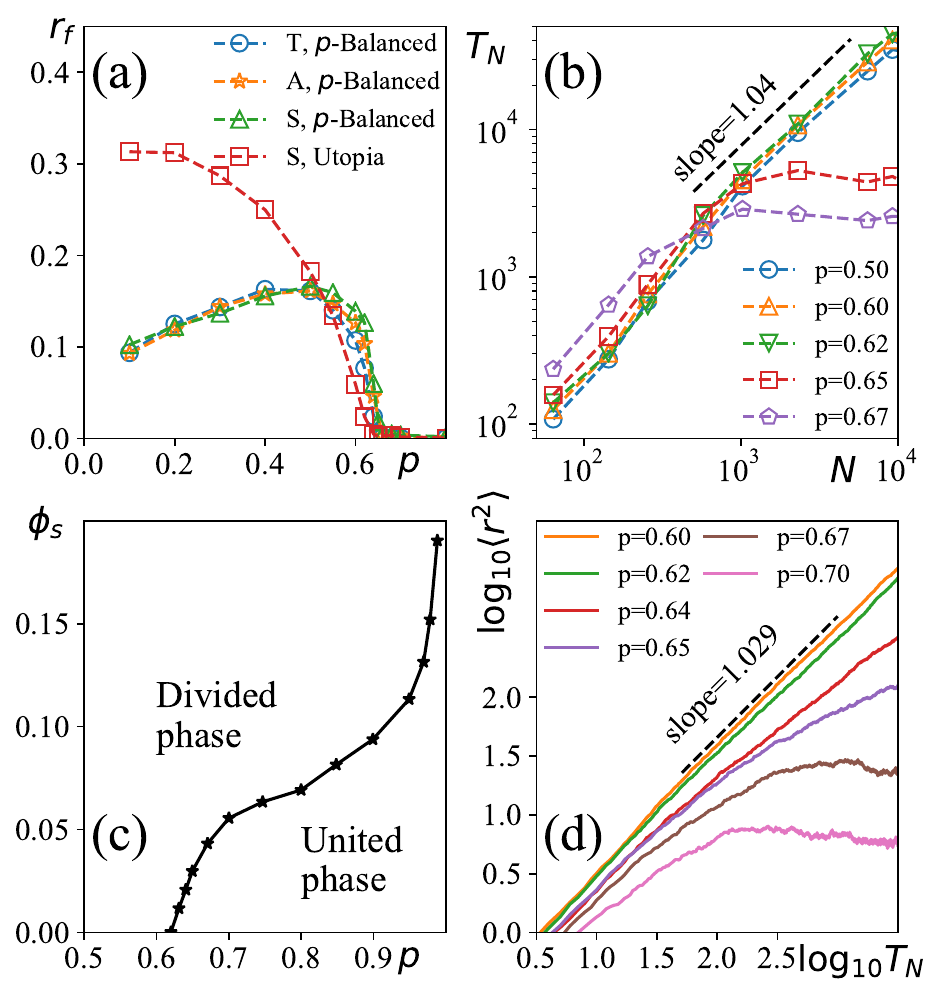}
	\vspace{-5mm}
	\caption{
	(a) Average ratio of flipped links v.s. propensity $p$, when a single rule-breaker is inserted. 
	(b) Evolution time scales approximately linearly with system size for $p >p_c$, and saturates to a finite limit for $p < p_c$. 	 
	(c) Phase diagram. {$\phi_S$} is the fraction of  separatists inserted into a utopia sate.  In the united phase, there is a system-spanning  clique, whereas in the divided phase, all cliques are of finite size.
	(d) Mean squared displacement (MSD) between two imbalanced triads as a function of time.  System size is $N=256\times 256$. }
	\label{fig:Fig2}
\vspace{-4mm}
\end{figure}

\vspace{1mm}
{\bf Rule-breakers} \quad  We insert a single rule-breaker node into a $p$-balanced society, and restart the $p$-dynamics. The insertion generally introduces one or more pairs of imbalanced triad~\footnote{The actual number of imbalanced triads produced by the insertion depends not only on the type of rule-breakers, but also on the microscopic state of the social network and the precise location of insertion.}, which move randomly and annihilate in pairs, until the society is balanced.  The percentage of all flipped links $r$ as a function of $p$ is plotted in Fig.~\ref{fig:Fig2}(a).  For $p > p_c\approx 0.64$ (the united phase), $r$ is negligible, whereas for $p >p_c$ (the divided phase), $r$ is finite, which means that a macroscopic rearrangement of the society is caused by the insertion.   We also compute the average time from insertion to re-balancing, as a function of system size.  As shown in Fig.~\ref{fig:Fig2}(b), the scaling is power-law with exponent $ 1.04$ for $p >p_c$.  For $p < p_c$, the evolution seems saturating to a finite limit as $N$ increases.  These scalings are of course consistent with the results shown in Fig.~\ref{fig:Fig2}(a).  Note however, these scalings are different from the scaling of the evolution time from a random imbalanced state to a $p$-balanced state as studied in Sec. I of SI.  See also Ref.~\cite{radicchi2007universality} for an earlier study of the latter issue.  





We now insert a rule-breaker into the utopia state, and restart the $p$-dynamics.  As shown in Fig.~\ref{fig::Fig1}(b),
 insertion of an angel leads to no change of the utopia state at all.  Interestingly, insertion of a thug only leads to flipping of six neighboring links, but the dynamics stops right away, since there is no imbalanced triad.  By strong contrast,  insertion of a separatist $S_m$ generates $m$ imbalanced triads, as illustrated in Fig.~\ref{fig::Fig1}(b).  As the $p$-dynamics is turned on, these imbalanced triads move randomly, before they annihilate.  The average percentage of flipped links $r$ before re-balancing is again negligible for $p > p_c$ but finite for $p < p_c$, as shown by the green curve in Fig.~\ref{fig:Fig2}(a).  For $p < p_c$, the final society is fragmented into many cliques of finite sizes that are separated by foe links.  Here we define a {\em clique} as a maximal cluster of nodes connected only by friend links, such that deletion of any single friend link does not disconnect the cluster.  A sample of fragmented society is shown in Fig.~\ref{fig::Fig1}(c).   We now insert a finite fraction of separatists in the utopia state, at randomly chosen positions.   The final balanced society, as shown in Fig.~\ref{fig:Fig2}(c), again exhibits a united phase with is a system-spanning clique, and a divided phase where the society is fragmented into many finite size cliques.  In the divided phase, insertion of a single separatist is sufficient to induce fragmentation of the entire network.  Hence, the utopia state with $p< p_c$ is intrinsically unstable.  

To understand the difference between the united phase and the divided phase, we generate two imbalanced triads by inserting a separatist $S_2$ into a $p$-balanced society~\footnote{If the number of generated imbalanced triads not two, we simply try it again.}, and prohibit annihilation, so that these imbalanced triads move indefinitely.  We plot the mean squared displacement (MSD) between these two triads as a function of time.  As shown in Fig.~\ref{fig:Fig2}(d), for $p < p_c$, we find MSD $\sim t$, corresponding to normal diffusion, which suggests that imbalanced triads behave as free particles with negligible interaction.   By strong contrast, for $p > p_c$, MSD saturates to a finite limit, indicating that the imbalanced triads are confined by long range attraction.  If annihilation were allowed, two imbalanced triads would quickly annihilate each other.   These results suggest that the united-divided phase transition on social network may be understood as a confinement-deconfinement transition of social tension.

	\vspace{1mm}
{\bf Ising lattice gauge theory}\quad It is remarkable that social tension, i.e. imbalanced triads, can be described as  defects in Ising lattice gauge theory~\cite{kogut1979introduction}.  This connection has been mysteriously overlooked to date.   Consider the Hamiltonian of an Ising gauge model on triangular lattice:
\ba
\beta H_{\rm IG} [ \sigma ] = -J\sum_{i,j,k \in \triangle} 
 \sigma_{ij} \sigma_{jk} \sigma_{ki}
 - h\sum_{\langle ij \rangle} \sigma_{ij}, \label{Hamiltonian-triang}
\vspace{-2mm}
\ea
where spin variables $\sigma_{ij}$ are defined on links, and $J, h$ are respectively dimensionless coupling constant and magnetic field.  The first term of $\beta H_{\rm IG} [ \sigma ]$ penalizes each imbalanced triad by energy $2J$, whilst the second term penalizes each foe link by energy $2h$.   It is easy to see that the social $p$-dynamics we studied above satisfies detailed balance with respect to the Gibbs distribution of Eq.~(\ref{Hamiltonian-triang}), if we choose $J = \infty$ and 
\ba
e^{2 h} = p/(1-p).  \label{h-p-corresp}
\vspace{-2mm}
\ea
Hence the signed social dynamics is precisely the equilibrium Monte-Carlo dynamics of Ising gauge model (\ref{Hamiltonian-triang}) in the limit of large $J$.



\begin{figure}[t!]
	\centering
	\includegraphics[width=3.4in]{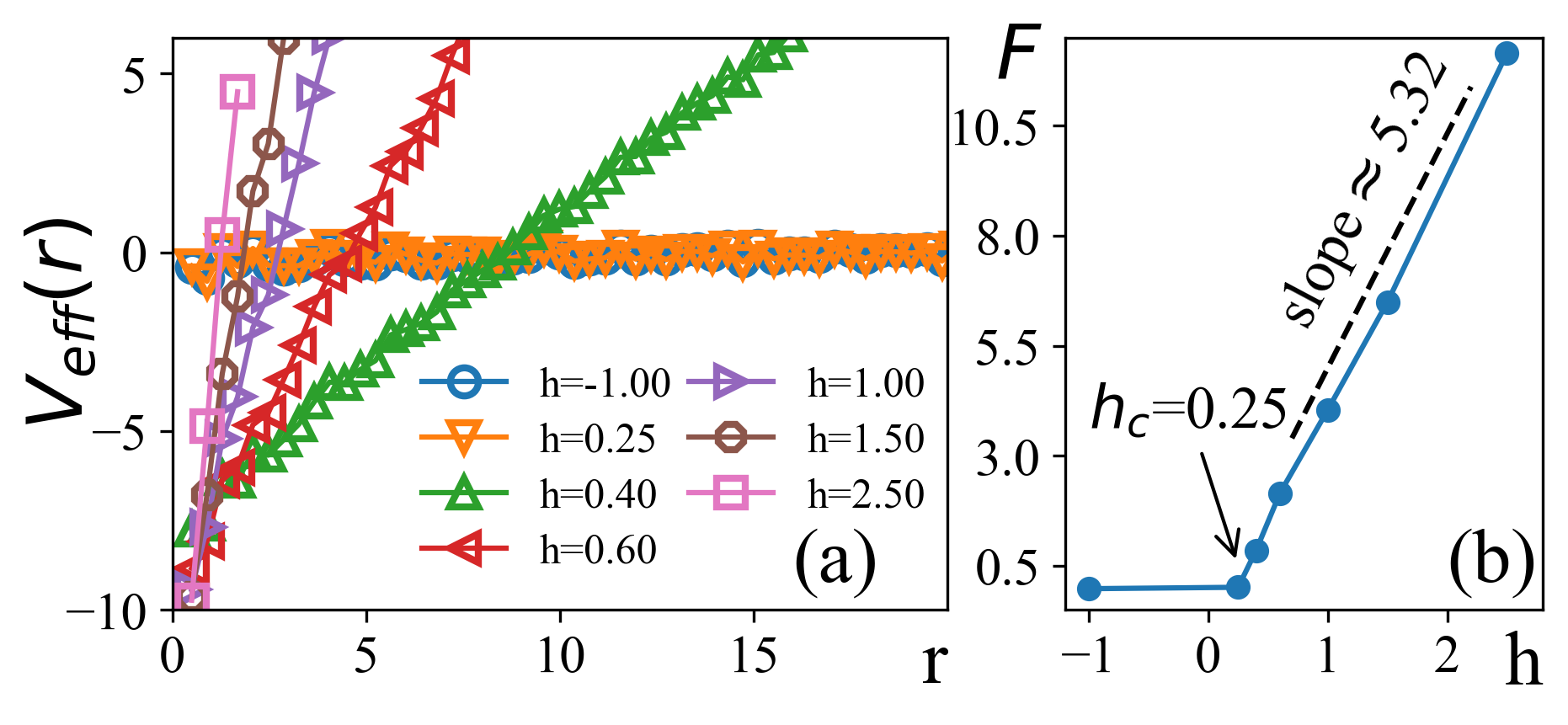}
	\vspace{-7mm}
	\caption{(a) Effectvie interaction and (b) force between visons.}
	\label{fig::effective_force_combine}
\vspace{-4mm}
\end{figure}




Using the language of Ising lattice gauge theory, an imbalanced triad, i.e., social tension,  behaves as a $Z_2$ topological defects, each carrying an energy $2 J$.  Following terminologies used in strongly correlated electron systems, we shall call these defects ``visons''~\cite{Senthil-2000-vison}.  Two visons may annihilate each other, resulting in decrease of energy by $4J$.  Fluctuations of spins lead to affective interaction between visons, which is given by the effective free energy of the spin system with positions of visons fixed.  We numerically compute this effective interaction between two visons using Monte-Carlo simulation.   The details are explained in Sec. II of SI.    As shown in Fig.~\ref{fig::effective_force_combine}(a), for $p > p_c$, the effective interaction between two visons increases linearly with distance, which means that these defects are linearly confined.  For $p < p_c$, the interaction is nearly independent of distance, which means that visons are deconfined.  In Fig.~\ref{fig::effective_force_combine}(b) we plot the slope of linear confining potential, i.e. the force between visons, as a function of field $h$.  It is seen that the slope vanishes at the critical field $h_c \approx 0.25$, and increases linearly with $h$ for $h > h_c$, suggesting that the transition is continuous.  Hence the physical mechanism of  united v.s. divided transition of social network is the deconfinement of social tension, i.e. $Z_2$ defects in the Ising lattice gauge theory.  This transition is of topological nature, because two phases are distinguished not by any local order parameter, but by the effective interaction between $Z_2$ defects.   


\begin{table}[h!]
\setlength{\extrarowheight}{2pt}
\begin{tabular}{ r | c | l }
\hline Social Network  \hspace{2mm}& United phase & \hspace{2mm}Divided phase \\ \hline
Ising Gauge Model   \hspace{2mm}& Confined  & \hspace{2mm}Deconfined  \\ \hline
Dual Ising Model  \hspace{.5mm} & \hspace{2mm} Ferromagnetic \hspace{2mm} & \hspace{1mm} Paramagnetic \\  \hline  
      \end{tabular}
\caption{Correspondence between social network model, Ising lattice gauge theory, and dual Ising model.}
      \label{table-I}
\end{table}

Ising gauge theory as defined by Eq.~(\ref{Hamiltonian-triang}) in two dimension does not have any phase transition for finite values of $J$ and $h$~\footnote{A proof for square lattice is given in Ref.~\cite{kogut1979introduction}.  It is not difficult to generalize the proof to triangular lattice. }.  Our social network model however corresponds to the limit $ J \rightarrow \infty$.  Within this limit, the configuration space of Eq.~(\ref{Hamiltonian-triang})  can be decomposed into many well separated  subspaces, each with $N_v = 0,2,4, \cdots$ visons, with the lowest subspace consisting of all balanced states.   According to Harary's theorem~\cite{harary1953notion,cartwright1956structural}, if the system is balanced, all nodes can be decomposed into two subsets, such that nodes within the same subset are connected only by friend links, whereas nodes from different subsets are connected only by foe links.  We can then assign Ising variables $s_i = 1$ to all nodes in the first subset, and $s_i = -1$ to all sites in the second subset, which yields $\sigma_{ij} = s_i s_j$, for all $i, j$.  This maps  the Hamiltonian (\ref{Hamiltonian-triang}) into the Hamiltonian of an Ising model on triangular lattice: $\beta H_{\rm I}[s] =  - h \sum_{\langle ij \rangle} s_i s_j,$
where the magnetic field $h$ now plays the role of coupling constant!  Now Ising model on triangular lattice does have a critical point $h_c = 0.2746...$~\cite{Domb-2000-book}, which separates a ferromagnetic phase ($h> h_c$) from a paramagnetic phase ($h < h_c$).  The corresponding critical propensity is determined using Eq.~(\ref{h-p-corresp}) as $p_c = 1/(1+ e^{- h_c}) \approx 0.634$, which is very close to our numerical result. The correspondence between the social network model, the Ising lattice gauge model, and the dual Ising model is shown in Table \ref{table-I}.

%

\vspace{2mm}

{\bf Social dynamics on other networks} \quad It is important to note that both social dynamics and Ising gauge theory can be studied on other regular lattices, or even on random networks, with proper adaptation of model details.  On 2d square lattice, for example, the role of triad balance should be replaced by square balance, i.e., the product of all four link variables on the same square must be $+1$.  In fact, Ising lattice gauge theory was mostly studied on square lattice and its higher dimensional analogues~\cite{kogut1979introduction,balian1975gauge}.   We have studied the same social dynamics as well as the mapping to Ising  gauge theory on square lattice.  The results are completely parallel to those on triangular lattice, as shown in Sec. III of SI.  In higher dimensional lattices, an important difference arises, because Ising lattice gauge theory has a confinement-deconfinement transition even for finite $J$.  This implies that even if imbalanced triads are allowed to appear spontaneously in the social dynamics, the social tension may still be confined as long as the propensity is high enough.  In this case, in the divided phase, social tension diffuses through the entire system as chages moving in a plasma, whereas in the united phase, they are bonded into pairs behave as neutral molecules.  

Real social networks are however better modeled as random networks.  We have also simulated the social dynamics on three random networks, starting from the utopia state and introducing one single separatist.  As shown in detail in Sec. IV of SI, for each system, there  exists a well-defined threshold propensity $p_c$ separating a united regime and a divided regime.  For $p > p_c$, less than half of the edges  are flipped, whereas for $p < p_c$, the majority of edges are flipped. For random networks of finite size, of course, we do not expect a sharp phase phase transition as in regular lattice.  However, the confinement v.s. deconfinement of $Z_2$ defects we discovered above is still much valuable in understanding of the marked difference between a united society and a divided society.

\vspace{2mm}
{\rm {\bf Conclusions} \quad  Social tension on a social network corresponds to $Z_2$ defects in Ising lattice gauge theory, which are deconfined in the divided phase ($p < p_c$), and confined in the united phase ($p>p_c$).  Insertion of a single rule-breaker in the divided phase leads to macroscopic rearrangement of social relations.  This connection between social dynamics and topological phase transition, which has been miraculously missed so far, worths more systematic exploration in the future.  On the other hand, even though macroscopic rearrangement of social relations is triggered by insertion of rule-breakers, the dynamics of rearrangement is actually driven by interactions between all constituents of the society.  This  interplay between ``rule-breakers'' and ``the silent majority'' provides an interesting perspective for reconciliation between the great man theory and the theory of ``history from below''.  



X.X. acknowllinks support from NSFC via grant \#11674217, as well as additional support from a Shanghai Talent Program.  This research is also supported by Shanghai Municipal Science and Technology Major Project (Grant No.2019SHZDZX01).


\begin{thebibliography}{10}
	
	\bibitem{Castellano-RMP-2009}
	Castellano, Claudio, Santo Fortunato, and Vittorio Loreto. 
	\newblock ``Statistical physics of social dynamics." 
	\newblock Reviews of modern physics 81.2 (2009): 591.
	
	\bibitem{Xia-Opinion-review-2011}
	Xia, Haoxiang, Huili Wang, and Zhaoguo Xuan. 
	\newblock ``Opinion dynamics: A multidisciplinary review and perspective on future 	research." 
	\newblock International Journal of Knowllink and Systems Science (IJKSS) 2.4 (2011): 72-91.

	\bibitem{Lorenz-Opinion-2007}
	Lorenz, Jan. 
	\newblock ``Continuous opinion dynamics under bounded confidence: A survey." 
	\newblock International Journal of Modern Physics C 18.12 (2007): 1819-1838.
	
	\bibitem{Martins-Opinion-2008}
	Martins, André CR. 
	\newblock ``Continuous opinions and discrete actions in opinion dynamics problems." 
	\newblock International Journal of Modern Physics C 19.04 (2008): 617-624.
	
	\bibitem{Wang-epidemic-2017}
	\newblock Wang, Wei, et al. 
	\newblock ``Unification of theoretical approaches for epidemic spreading on complex networks." 
	\newblock Reports on Progress in Physics 80.3 (2017): 036603.
	
	\bibitem{Pastor-epidemic-2001}
	\newblock Pastor-Satorras, Romualdo, and Alessandro Vespignani. 
	\newblock ``Epidemic spreading in scale-free networks." Physical review letters 86.14 (2001): 3200.

	\bibitem{Castellano-epidemic-2010}
	Castellano, Claudio, and Romualdo Pastor-Satorras. 
	\newblock ``Thresholds for epidemic spreading in networks." 
	\newblock Physical review letters 105.21 (2010): 218701.

	\bibitem{Axelrod-culture-1997}
	Axelrod, Robert. 
	\newblock ``The dissemination of culture: A model with local convergence and global polarization." 
	\newblock Journal of conflict resolution 41.2 (1997): 203-226.
	
	\bibitem{Castellano-culture-2000}
	Castellano, Claudio, Matteo Marsili, and Alessandro Vespignani. 
	\newblock ``Nonequilibrium phase transition in a model for social influence." 
	\newblock Physical Review Letters 85.16 (2000): 3536.

	\bibitem{Abrams-language-2003}
	Abrams, Daniel M., and Steven H. Strogatz. 
	\newblock ``Modelling the dynamics of language death." 
	\newblock Nature 424.6951 (2003): 900-900.

	\bibitem{Vicsek-crowd-1995}
	Vicsek, Tamás, et al. 
	\newblock ``Novel type of phase transition in a system of self-driven particles." 
	\newblock Physical review letters 75.6 (1995): 1226.

	\bibitem{Strogatz-crowd-1993}
	\newblock ``Strogatz, Steven H., and Ian Stewart. "Coupled oscillators and biological synchronization." 
	\newblock Scientific American 269.6 (1993): 102-109.

	\bibitem{Bonabeau-hierarchies-1995}
	Bonabeau, Eric, Guy Theraulaz, and Jean-Louis Deneubourg. 
	\newblock ``Phase diagram of a model of self-organizing hierarchies." 
	\newblock Physica A: Statistical Mechanics and its Applications 217.3-4 (1995): 373-392.

	\bibitem{Ben-Naim-hierarchies-2005}
	Ben-Naim, E., and S. Redner. 
	\newblock ``Dynamics of social diversity." 
	\newblock  Journal of Statistical Mechanics: Theory and Experiment 2005.11 (2005): L11002.

	\bibitem{Carlyle-hero-1869}
	Carlyle, Thomas. 
	\newblock Heroes and hero-worship. Vol. 12. 
	\newblock Chapman and Hall, 1869.

	\bibitem{Hook-hero-1943}
	Hook, Sidney. 
	\newblock The hero in history: A study in limitation and possibility. 
	\newblock Transaction Publishers, 1943.

	\bibitem{Segal-hero-2000}
	Segal, Robert. 
	\newblock Hero myths: A reader. 
	\newblock John Wiley \& Sons Ltd., 2000.

	\bibitem{Faulkner-hero-2008}
	Faulkner, Robert. 
	\newblock The case for greatness: Honorable ambition and its critics. 
	\newblock Yale University Press, 2008.

	\bibitem{Thompson-history-below-1966}
	Thompson, Edward Palmer. 
	\newblock ``History from below." 
	\newblock Times Literary Supplement 7.04 (1966): 76-106.

	\bibitem{heider1946attitudes}
	Fritz Heider.
	\newblock Attitudes and cognitive organization.
	\newblock {\em The Journal of psychology}, 21(1):107--112, 1946.
	
	\bibitem{harary1953notion}
	Frank Harary et~al.
	\newblock On the notion of balance of a signed graph.
	\newblock {\em Michigan Mathematical Journal}, 2(2):143--146, 1953.
	
	\bibitem{cartwright1956structural}
	Dorwin Cartwright and Frank Harary.
	\newblock Structural balance: a generalization of heider's theory.
	\newblock {\em Psychological review}, 63(5):277, 1956.
	
	\bibitem{wasserman1994social}
	Stanley Wasserman, Katherine Faust, et~al.
	\newblock Social network analysis: Methods and applications.
	\newblock 1994.
	
	\bibitem{zheng2015social}
	Xiaolong Zheng, Daniel Zeng, and Fei-Yue Wang.
	\newblock Social balance in signed networks.
	\newblock {\em Information Systems Frontiers}, 17(5):1077--1095, 2015.
	
	\bibitem{leskovec2010predicting}
	Jure Leskovec, Daniel Huttenlocher, and Jon Kleinberg.
	\newblock Predicting positive and negative links in online social networks.
	\newblock In {\em Proceedings of the 19th international conference on World
		wide web}, pages 641--650, 2010.
\\	
	Liben‐Nowell, David, and Jon Kleinberg. 
	\newblock The link‐prediction problem for social networks. 
	\newblock {\em Journal of the American society for information science and technology} 58.7 (2007): 1019-1031.
	
	\bibitem{facchetti2011computing}
	Giuseppe Facchetti, Giovanni Iacono, and Claudio Altafini.
	\newblock Computing global structural balance in large-scale signed social
	networks.
	\newblock {\em Proceedings of the National Academy of Sciences},
	108(52):20953--20958, 2011.
	
	\bibitem{aref2018measuring}
	Samin Aref and Mark~C Wilson.
	\newblock Measuring partial balance in signed networks.
	\newblock {\em Journal of Complex Networks}, 6(4):566--595, 2018.
	
	\bibitem{kirkley2019balance}
	Alec Kirkley, George~T Cantwell, and MEJ Newman.
	\newblock Balance in signed networks.
	\newblock {\em Physical Review E}, 99(1):012320, 2019.
	
	\bibitem{jung2016personalized}
	Jinhong Jung, Woojeong Jin, Lee Sael, and U~Kang.
	\newblock Personalized ranking in signed networks using signed random walk with
	restart.
	\newblock In {\em 2016 IEEE 16th International Conference on Data Mining
		(ICDM)}, pages 973--978. IEEE, 2016. 
	
	\bibitem{newman2006modularity}
	Mark~EJ Newman. 
	\newblock Modularity and community structure in networks.
	\newblock {\em Proceedings of the national academy of sciences},
	103(23):8577--8582, 2006.
	
	\bibitem{esmailian2015community}
	Pouya Esmailian and Mahdi Jalili.
	\newblock {Community detection} in signed networks: the role of negative ties in
	different scales.
	\newblock {\em Scientific reports}, 5(1):1--17, 2015.
	
	\bibitem{girdhar2019community}
	Nancy Girdhar and Kamal~Kant Bharadwaj.
	\newblock { Community detection} in signed social networks using multiobjective
	genetic algorithm.
	\newblock {\em Journal of the Association for Information Science and
		Technology}, 70(8):788--804, 2019.
	
	\bibitem{axelrod1993landscape}
	Robert Axelrod and D~Scott Bennett.
	\newblock A landscape theory of aggregation.
	\newblock {\em British journal of political science}, pages 211--233, 1993.
	
	\bibitem{altafini2012consensus}
	Claudio Altafini.
	\newblock Consensus problems on networks with antagonistic interactions.
	\newblock {\em IEEE Transactions on Automatic Control}, 58(4):935--946, 2012.
	
	\bibitem{altafini2012dynamics}
	Claudio Altafini.
	\newblock Dynamics of opinion forming in structurally balanced social networks.
	\newblock {\em PloS one}, 7(6):e38135, 2012.
	
	\bibitem{proskurnikov2017tutorial}
	Anton~V Proskurnikov and Roberto Tempo.
	\newblock A tutorial on modeling and analysis of dynamic social networks. part i.
	\newblock {\em Annual Reviews in Control}, 43:65--79, 2017.
	
	\bibitem{proskurnikov2018tutorial}
	Anton~V Proskurnikov and Roberto Tempo.
	\newblock A tutorial on modeling and analysis of dynamic social networks. part ii.
	\newblock {\em Annual Reviews in Control}, 45:166--190, 2018.

	\bibitem{hummon2003some}
	Norman~P Hummon and Patrick Doreian.
	\newblock Some dynamics of social balance processes: bringing Heider back into
	balance theory.
	\newblock {\em Social Networks}, 25(1):17--49, 2003.



	
	\bibitem{antal2005dynamics}
	Tibor Antal, Pavel~L Krapivsky, and Sidney Redner.
	\newblock Dynamics of social balance on networks.
	\newblock {\em Physical Review E}, 72(3):036121, 2005.
	
	
	\bibitem{antal2006social}
	Tibor Antal, Paul~L Krapivsky, and Sidney Redner.
	\newblock Social balance on networks: The dynamics of friendship and enmity.
	\newblock {\em Physica D: Nonlinear Phenomena}, 224(1-2):130--136, 2006.
	
	\bibitem{radicchi2007social}
	Filippo Radicchi, Daniele Vilone, Sooeyon Yoon, and Hildegard Meyer-Ortmanns.
	\newblock Social balance as a satisfiability problem of computer science.
	\newblock {\em Physical Review E}, 75(2):026106, 2007.
	
	\bibitem{radicchi2007universality}
	Filippo Radicchi, Daniele Vilone, and Hildegard Meyer-Ortmanns.
	\newblock Universality class of triad dynamics on a triangular lattice.
	\newblock {\em Physical Review E}, 75(2):021118, 2007.
	
	\bibitem{marvel2009energy}
	Seth~A Marvel, Steven~H Strogatz, and Jon~M Kleinberg.
	\newblock Energy landscape of social balance.
	\newblock {\em Physical review letters}, 103(19):198701, 2009.
	
	\bibitem{marvel2011continuous}
	Seth~A Marvel, Jon Kleinberg, Robert~D Kleinberg, and Steven~H Strogatz.
	\newblock Continuous-time model of structural balance.
	\newblock {\em Proceedings of the National Academy of Sciences},
	108(5):1771--1776, 2011.
	
	\bibitem{traag2013dynamical}
	Vincent~Antonio Traag, Paul Van~Dooren, and Patrick De~Leenheer.
	\newblock Dynamical models explaining social balance and evolution of
	cooperation.
	\newblock {\em PloS one}, 8(4):e60063, 2013.
	
	\bibitem{macy2003polarization}
	Michael~W Macy, James~A Kitts, Andreas Flache, and Steve Benard.
	\newblock Polarization in dynamic networks: A hopfield model of emergent
	structure.
	\newblock 2003.
	
	\bibitem{gross2019rise}
	J{\"o}rg Gross and Carsten~KW De~Dreu.
	\newblock The rise and fall of cooperation through reputation and group
	polarization.
	\newblock {\em Nature communications}, 10(1):1--10, 2019.
	
	\bibitem{saeedian2019absorbing}
	Meghdad Saeedian, Maxi San~Miguel, and Raul Toral.
	\newblock Absorbing phase transition in the coupled dynamics of node and link
	states in random networks.
	\newblock {\em Scientific reports}, 9(1):1--14, 2019.
	
	\bibitem{pham2020effect}
	Tuan~Minh Pham, Imre Kondor, Rudolf Hanel, and Stefan Thurner.
	\newblock The effect of social balance on social fragmentation.
	\newblock {\em Journal of the Royal Society Interface}, 17(172), 2020.
	
	\bibitem{gorski2020homophily}
	Piotr~J G{\'o}rski, Klavdiya Bochenina, Janusz~A Ho{\l}yst, and Raissa~M
	D’Souza.
	\newblock Homophily based on few attributes can impede structural balance.
	\newblock {\em Physical Review Letters}, 125(7):078302, 2020.
	
	\bibitem{Pham2021}
	Tuan~M. Pham, Andrew~C. Alexander, Jan Korbel, Rudolf Hanel, and Stefan
	Thurner.
	\newblock Balance and fragmentation in societies with homophily and social
	balance.
	\newblock {\em Scientific Reports}, 11(1):17188, 2021.
	
	\bibitem{heider2013psychology}
	Fritz Heider.
	\newblock {\em The psychology of interpersonal relations}.
	\newblock Psychology Press, 2013.
	
%
%
%
	
	\bibitem{kogut1979introduction}
	John~B Kogut.
	\newblock An introduction to lattice gauge theory and spin systems.
	\newblock {\em Reviews of Modern Physics}, 51(4):659, 1979.
	
	\bibitem{balian1975gauge}
	R~Balian, JM~Drouffe, and Claude Itzykson.
	\newblock Gauge fields on a lattice. ii. gauge-invariant ising model.
	\newblock {\em Physical Review D}, 11(8):2098, 1975.
	
	\bibitem{Domb-2000-book}	
	Domb, Cyril. 
	\newblock Phase transitions and critical phenomena. 
	\newblock Elsevier, 2000.
	
%
	
	\bibitem{barabasi1999emergence}
	Albert-L{\'a}szl{\'o} Barab{\'a}si and R{\'e}ka Albert.
	\newblock Emergence of scaling in random networks.
	\newblock {\em science}, 286(5439):509--512, 1999.
	
	\bibitem{watts1998collective}
	Duncan~J Watts and Steven~H Strogatz.
	\newblock Collective dynamics of ‘small-world’networks.
	\newblock {\em nature}, 393(6684):440--442, 1998.
	
	\bibitem{read1954cultures}
	Kenneth~E Read.
	\newblock Cultures of the central highlands, new guinea.
	\newblock {\em Southwestern Journal of Anthropology}, 10(1):1--43, 1954.
	
	\bibitem{nr}
	Ryan~A. Rossi and Nesreen~K. Ahmed.
	\newblock The network data repository with interactive graph analytics and
	visualization.
	\newblock In {\em AAAI}, 2015.
	
%
	
	\bibitem{Senthil-2000-vison}
	Senthil, T., and Matthew PA Fisher. 
	\newblock ``Z 2 gauge theory of electron fractionalization in strongly correlated systems."
	\newblock Physical Review B 62.12 (2000): 7850.
	
\end{thebibliography}

\end{document}